\documentclass[conference]{IEEEtran}
\IEEEoverridecommandlockouts
\usepackage{cite}
\usepackage{amsmath,amssymb,amsfonts}
\usepackage{algorithmic}
\usepackage{graphicx}
\usepackage{textcomp}
\usepackage{xcolor}
\usepackage[capitalise]{cleveref}

\usepackage{fp,tikz,pgfplots}
\usetikzlibrary{arrows,shapes,backgrounds,patterns,fadings,matrix,arrows,calc,
	intersections,decorations.markings,
	positioning,arrows.meta}
\usepgfplotslibrary{fillbetween}
\usepgfplotslibrary{statistics}
\pgfplotsset{width=5\columnwidth /5, compat = 1.13,
	height = 60\columnwidth /100, grid= major,
	legend cell align = left, ticklabel style = {font=\scriptsize},
	every axis label/.append style={font=\small},
	legend style = {font={\scriptsize}},title style={yshift=-7pt, font = \small} }

\newtheorem{remark}{Remark}

\xdefinecolor{bluePlot}{RGB}{129,155,255}
\xdefinecolor{greenPlot}{RGB}{154,255,147}
\xdefinecolor{yellowPlot}{RGB}{254,217,166}
\xdefinecolor{pinkPlot}{RGB}{255,105,50}

\xdefinecolor{blackPlot}{RGB}{100,100,100}
\xdefinecolor{greenPlot}{RGB}{102,194,165}
\xdefinecolor{bluePlot}{RGB}{141,160,203}
\xdefinecolor{redPlot}{RGB}{252,141,98}
\xdefinecolor{dataPoint1}{RGB}{166,206,227}
\xdefinecolor{dataPoint2}{RGB}{31,120,180}
\xdefinecolor{dataPoint3}{RGB}{178,223,138}
\xdefinecolor{dataPoint4}{RGB}{51,160,44}

\usepackage[normalem]{ulem}

\def\BibTeX{{\rm B\kern-.05em{\sc i\kern-.025em b}\kern-.08em
    T\kern-.1667em\lower.7ex\hbox{E}\kern-.125emX}}
\begin{document}

\title{Learning-based Control for PMSM Using Distributed Gaussian Processes with Optimal Aggregation Strategy
\thanks{$\dagger$  Equal  Contribution.\quad *Corresponding Author}
\thanks{This work was supported by a grant (Project No. 2022A1515110361) from Guangdong Basic and Applied Basic Research Foundation, China, and the Federal Ministry of Education and Research of Germany in the programme of “Souverän. Digital. Vernetzt.”. Joint project 6G-life, project identification number: 16KISK002.}
}

\author{\IEEEauthorblockN{1\textsuperscript{st} Zhenxiao Yin$^{\dagger }$}
\IEEEauthorblockA{\textit{Robotics and Autonomous Systems} \\
\textit{Thrust, The Hong Kong University of }\\
\textit{ Science and Technology (Guangzhou)} \\
Guangzhou, China \\
zyin368@connect.hkust-gz.edu.cn}
\and
\IEEEauthorblockN{1\textsuperscript{st} Xiaobing Dai$^{\dagger} $}
\IEEEauthorblockA{\textit{Chair of Information-oriented Control} \\
\textit{Technical University of Munich}\\
Munich, Germany \\
xiaobing.dai@tum.de}
\and
\IEEEauthorblockN{1\textsuperscript{st} Zewen Yang$^{\dagger} $}
\IEEEauthorblockA{\textit{Center for Artificial Intelligence} \\
\textit{in Public Health Research (ZKI-PH)} \\
\textit{Robert Koch Institute}\\
Berlin, Germany \\
yangz@rki.de}
\and
\IEEEauthorblockN{4\textsuperscript{th} Yang Shen}
\IEEEauthorblockA{\textit{Robotics and Autonomous Systems} \\
\textit{Thrust, The Hong Kong University of}\\
\textit{ Science and Technology (Guangzhou)} \\
Guangzhou, China \\
yangshen@hkust-gz.edu.cn}
\and
\IEEEauthorblockN{5\textsuperscript{th} Georges Hattab}
\IEEEauthorblockA{\textit{Center for Artificial Intelligence} \\
\textit{in Public Health Research (ZKI-PH)} \\
\textit{Robert Koch Institute}\\
Berlin, Germany \\
\textit{Department of Mathematics and Computer }\\
\textit{Science, Freie Universität Berlin}\\
Berlin, Germany \\
hattabg@rki.de}
\and
\IEEEauthorblockN{6\textsuperscript{th} Hang Zhao$^*$}
\IEEEauthorblockA{
\textit{Robotics and Autonomous Systems} \\
\textit{Thrust, The Hong Kong University of}\\
\textit{Science and Technology (Guangzhou)} \\
Guangzhou, China \\
\textit{Department of Electronic \& Computer}\\
\textit{Engineering, The Hong Kong}\\
\textit{University of Science and Technology}\\
Hong Kong, China \\
hangzhao@ust.hk}}
\maketitle
\vspace{-0.3cm}
\begin{abstract}
The growing demand for accurate control in varying and unknown environments has sparked a corresponding increase in the requirements for power supply components, including permanent magnet synchronous motors (PMSMs). To infer the unknown part of the system, machine learning techniques are widely employed, especially Gaussian process regression (GPR) due to its flexibility of continuous system modeling and its guaranteed performance. For practical implementation, distributed GPR is adopted to alleviate the high computational complexity. However, the study of distributed GPR from a control perspective remains an open problem. In this paper, a control-aware optimal aggregation strategy of distributed GPR for PMSMs is proposed based on the Lyapunov stability theory. This strategy exclusively leverages the posterior mean, thereby obviating the need for computationally intensive calculations associated with posterior variance in alternative approaches. Moreover, the straightforward calculation process of our proposed strategy lends itself to seamless implementation in high-frequency PMSM control. The effectiveness of the proposed strategy is demonstrated in the simulations. \looseness=-1
\end{abstract}


\section{Introduction}
In recent years, permanent magnet synchronous motors (PMSMs), serving as crucial components and energy sources, find widespread applications in diverse industries, including electric vehicles~\cite{zhang2021coupling}, underwater vehicles~\cite{yan2018novel} and soft robotics~\cite{gao2023quasi}. Driven by the increasing requirements on adaptivity and robustness, the demand for precise control of PMSMs under varying environmental conditions has been increasing.

Due to the intricate system model and inherent environmental uncertainties, achieving accurate control of PMSMs poses a significant challenge. 
One commonly adopted technique for addressing this challenge is robust control~\cite{zhou1998essentials}, such as the proportional-integral (PI) controller~\cite{apte2019disturbance}, which aims to keep the operating error bounded under unknown bounded external disturbances. In robust control, large control gains are always employed to reduce the effects of disturbances, but also increase the sensitivity on the accuracy of state measurements, inducing high requirements on the real sensors~\cite{bovskovic2020novel}. 
Forward compensation of unknown external disturbances is another practical approach for unknown system control~\cite{zhao2015adaptive}, where the unknown part is predicted in advance using data-driven techniques such as machine learning methods. 
For PMSM control, learning-based control methods, such as Neural Network (NN) control~\cite{yin2022implementation}, have been widely studied.
However, the quantitative of its prediction performance lacks, primarily due to its heavy reliance on trained features~\cite{giles1987learning}. 
As a result, this limitation hampers its practical application in safety-critical industries.
Gaussian Process Regression (GPR)~\cite{williams2006gaussian}, as a supervised Bayesian-based machine learning technique, draws increasing attention in safe control areas, such as for feedback linearization~\cite{dai2023can} and back-stepping~\cite{ignatyev2023sparse}. 
The wide application results from its modeling flexibility for nonlinear functions and the existence of the theoretical prediction error bound.
The prediction performance of GPR benefits from the high data density.
However, the computational complexity of GPR, also strongly related to the number of training samples, poses challenges to its practical implementation in real systems~\cite{dai2023can}.
To reduce the computation time for GP in real-time PMSM control, many methods are proposed, such as inducing point methods~\cite{cao2013efficient}, finite feature approximations~\cite{mutny2018efficient}, or distributed GPR models~\cite{deisenroth2015distributed}.
Inheriting the prediction error bound from the single GPR, distributed GPR is commonly employed in real-time safety-critical scenarios~\cite{lederer2022networked}, which separates the entire data set into several models.
To this end, the development of smart data set separation, such as K-means clustering~\cite{melton2020k} and locally growing trees~\cite{lederer2021gaussian}, and aggregation techniques become necessary. 
In this paper, the aggregation techniques for distributed GPR from a control perspective are studied. 

The aggregation structures for distributed GPR models are widely studied, such as the mixture of experts (MOE)~\cite{tresp2000mixtures}, generalized product of experts (GPOE)~\cite{cao2014generalized} and Bayesian committee machine (BCM)~\cite{tresp2000bayesian}. 
However, the selection strategy for aggregation weights is still spontaneous, only having suggestions with identical weights or using an information-theoretic method~\cite{liu2020gaussian}. 
Furthermore, the computation of the prediction variance is required in these methods, which is more time-consuming in GPR prediction, but has few effects on controllers. 
The LoG-GP technique offers an online data separation and aggregation strategy that relies solely on the current input, without the computation of the variance~\cite{lederer2021gaussian}. 
However, its applicability is limited to distributed data sets organized in a tree structure generated by its online learning algorithm, and cannot be applied to arbitrary structures of data sets. 
Additionally, while the aforementioned methods maintain the validity of the prediction error bound~\cite{srinivas2012information}, they are primarily designed to enhance prediction accuracy rather than address control performance. 
Consequently, there is an urgent need for a control-oriented aggregation strategy that can accommodate arbitrary data structures while minimizing computational requirements and maintaining performance guarantees.

In this paper, a control-aware optimal aggregation strategy for GPR-based control of PMSM is proposed, which uses only the posterior means from distributed GPR model, saving calculations for variance. The aggregation weights are derived through the constrained linear programming problem, which is explicitly solved without complex numerical methods. The low computation load from the posterior mean and optimization solution enables its practical implementation in the control hardware with high frequency. Moreover, the proposed strategy inherits the existence of theoretical error bound for vanilla GPR~\cite{srinivas2012information}, which is adopted to derive the guaranteed control performance. The effectiveness of the proposed fusion learning-based control is demonstrated in the simulation.

\section{Problem Setting and Preliminaries}

\subsection{Motion Dynamical Model of PMSM}
In this paper,  the motion dynamics of a PMSM model with the system state $x=[\varphi, \omega]^T$  is considered as follows

\begin{equation}\label{eq_system}
    \dot{x}= \frac{\mathrm{d} }{\mathrm{d} t} \begin{bmatrix}
        \varphi \\ \omega
    \end{bmatrix} = \begin{bmatrix}
        \omega \\ J^{-1}\left(-B \omega+1.5 p \psi I_{q}-T\right)
    \end{bmatrix},
\end{equation}
where $\varphi$ and $\omega$ denote the rotation angle and angular velocity, respectively. The $q$-axis current  $I_q$ is the control input, and $T$ is the external torque. The constant motor parameters are known including damping ratio $B$, pole pair $p$, flux linkage $\psi$, and moment of inertia $J$. 

Considering the limited speed capability, the angular velocity $\omega$ is assumed to be bounded, i.e., $\omega \in [ \underline{\omega},\bar{\omega}]$ . Moreover, the external torque $T$ usually depends on the working situation characterized by $\varphi$ and $\omega$, for instance for manipulators with bounded $\varphi$, or engines with periodic $\varphi$. However, due to the complex structure of the driven mechanism and external environmental uncertainties, the applied torque $T$ on the PMSM is usually unknown. Therefore, it is assumed that the mapping between the torque $T$ and $\varphi_m$ and $\omega$ obeys $T= f(x_m)$ with
\begin{equation} \label{eqn_x2xm}
    x_m = \begin{bmatrix}
        \varphi_m \\ \omega_m
    \end{bmatrix} = \begin{bmatrix}
        \operatorname{mod}(\upsilon \varphi, 2\pi) - \pi \\
        \omega/ \omega_{\max}
    \end{bmatrix},
\end{equation}
where $\operatorname{mod}(\cdot, \cdot)$ denotes the modulo calculation and $\upsilon$ represents the ratio between the output rotation of the PMSM and the input motion of the driven mechanism. The constant $\omega_{\max}$ is the maximal absolute value of the angular velocity, i.e., $\omega_{\max} = \max(| \underline{\omega}|, |\bar{\omega}|)$. Due to the modulo operator,  $\varphi_m$ is bounded within $[-\pi, \pi )$. Combined with the bounded angular velocity, the input domain of $f(\cdot)$ is covered by a compact domain $\mathbb{X}$ i.e., $ \mathbb{X} = [-\pi, \pi] \times [-1, 1]$.

For the safe operation of the driven mechanism, high accuracy in both rotational position and velocity are required. Therefore, the control task of tracking a continuous and at least second-order differentiable reference $\varphi_d$ is considered. This task also aligns with the speed control for PMSM, since the speed reference is easily derived as $\omega_d = \dot{\varphi}_d$  and the angle part is treated as its integration for PI control.

\subsection{Generation of Training Data}
For the inference of unknown function $f(\cdot)$, data-driven techniques, such as GPR, are employed. The training data set with $M$ training samples is defined as follows
\begin{equation}
    \mathbb{D}=\left\{x_{G P}^{(i)}, y^{(i)}=f\left(x^{(i)}\right)+\epsilon_{T}^{(i)}\right\}_{t=1}^{M},
\end{equation}
where the measurement noise $\epsilon_{T}^{(i)}$ satisfies an identical, independent, zero-mean Gaussian distribution with $\sigma_T >0$, i.e., $\epsilon_{T}^{(i)} \sim \mathcal{N}(0, \sigma_T^2)$. The conventional data-driven methods require noise-free measurement of Comparing with the inputs and outputs of $f(\cdot)$, the inputs $x_{GP}$ correspond to $x_m$, whose noise-free measurements are unobtainable due to imperfect sensors in reality. Moreover, the commonly used sensors, i.e., encoders, only detect the relative rotation of PMSM, i.e.,
\begin{equation}
    z_{k}=\varphi_{m, k}+\epsilon_{\varphi, k}, \quad \epsilon_{\varphi, k} \sim \mathcal{N}\left(0, \sigma_{\varphi}^{2}\right),
\end{equation}
where $\bullet_k := \bullet(t_k)$ for simplification and $\sigma_{\varphi}>0$ is the variance of measurement noise. The derivative of $\varphi_{m, k}$ is usually obtained through numerical methods. For example, using the finite difference, the computed angular velocity and acceleration are written as
\begin{align}
  \omega_{k}&=\tilde{\omega}_{k}-\epsilon_{\omega, k}=\frac{1}{\Delta} \left(z_{k+1}-z_{k}\right)-\epsilon_{\omega, k}, \\
\dot{\omega}_{k}&=\frac{1}{\Delta^2} \left( z_{k+2}-2 z_{k+1}+z_{k} \right) -\epsilon_{d \omega},  
\end{align}
where $\epsilon_{\omega, k} \sim \mathcal{N}\left(0, \sigma_{\omega}^{2}\right)$, $ \epsilon_{d \omega} \sim \mathcal{N}\left(0, \sigma_{d \omega}^{2}\right)$, $\sigma_{\omega}^{2}=4 \Delta^{-2} \sigma_{\varphi}^{2}+\sigma_{l i n, \varphi}^{2}$, and $  \sigma_{d \omega}^{2}=4 \Delta^{-2} \sigma_{\omega}^{2}+\sigma_{l i n, \omega}^{2}$. The symbol $\Delta$ represents the sampling interval, while $\sigma_{l i n, \varphi}$ and $\sigma_{l i n, \omega}$ denote the variances associated with the errors in linearizing angular velocity and acceleration, respectively. Then, the computed external torque is written as
\begin{align}
    T_{k} :&=\tilde{T}_{k}-\epsilon_{T, k} \nonumber\\
    & =-J \dot{\omega}_{k}-B \omega_{k}+1.5 p \psi I_{q, k} \nonumber\\
    & =J \Delta\left(\tilde{\omega}_{k+1}-\tilde{\omega}_{k}\right)-B \tilde{\omega}_{k}+1.5 p \psi I_{q, k}-\epsilon_{T, k},
\end{align}
where the variance of noise $\epsilon_T$ is derived as
\begin{equation}
    \sigma_{T}^{2}=\left(\frac{4 J^{2}}{\Delta^2} +B^{2}\right) \sigma_{\omega}^{2}+J^{2} \sigma_{lin, \omega}^{2}.
\end{equation}
With this method, the measurement and linearization errors are transmitted to $y = \tilde{T}$ with $x_{GP}$ defined as $x_{GP} = [z_k, \tilde{\omega}_k]^T$. Note that the distribution of the modified measurement error $\upsilon_T$ not only depends on the measurement accuracy of $\omega_m$, but also relies on the sampling rate reflected by $\Delta$ and linearization errors $\epsilon_{\omega, k}$ and $\epsilon_{d\omega, k}$. Smaller sampling interval will directly lead to large output noise but also indicates small linearization errors. The proper choice of $\Delta$ is important for obtaining accurate training data, but it is out of the scope of this paper and is left as future work.

\subsection{Gaussian Process Regression}
Gaussian process regression is a supervised learning technique employed to infer unknown functions using measurable data sets. It leverages the principles of probability and statistics to estimate these functions based on the provided data. The features of the unknown function are encoded into the mean function $m(x)$ and kernel function $\kappa(x, x')$, inducing a distribution over $f(\cdot)$, i.e., $f(\cdot)\sim \mathcal{GP}(m(x), \kappa(x, x'))$. The mean function collects the partially known part of $f(\cdot)$, which is usually set as $m(x)=0$ for a completely unknown function. Considering the continuity of $f(\cdot)$ with compact input domain $\mathbb{X}$, the kernel function $\kappa(\cdot, \cdot)$ is selected as stationary, monotonically decreasing and Lipschitz continuous, such as square exponential kernel

In a regression task, Gaussian processes employ the joint distribution of a training data set $\mathbb{D}$ and a query point $x^*$ as
\begin{equation}
    \begin{bmatrix}
 Y\\
f(x^*)
\end{bmatrix} \sim \mathcal{N}\left(\begin{bmatrix}
 0\\
0
\end{bmatrix},\begin{bmatrix}
K+\sigma_T^2 I_M  & \kappa_{X}\\
\kappa_X^T  & \kappa(x^*,x^*)
\end{bmatrix} \right),
\end{equation}
where $Y = [y^{(1)}, \cdots, y^{(M)}]^T $, $K = \left[ \kappa(x^{(i)}, x^{(j)}) \right]_{i,j = 1, \cdots M}$, and $\kappa_X \!=\! [\kappa(x^*,x^{(1)}, \!\cdots\!, \kappa(x^*,x^{(M)})]^T$. Applying Bayes' theory, the posterior mean and variance at the test point $x^*$ are obtained through
\begin{align}\label{eq_prediction}
    \mu(x^*, \mathbb{D}) &= \kappa_X^T \left (  K+ \sigma_T^2 I\right )^{-1}  Y, \\
    \sigma(x^*, \mathbb{D}) &= \kappa(x^*,x^*) - \kappa_X^T \left (  K+ \sigma_T^2 I\right )^{-1} \kappa_X,
\end{align}
respectively. Using the properties of Gaussian distribution and Lipschitz continuity, and picking $\delta \in (0,1)$ and $\tau>0$, the prediction error from Gaussian process regression is uniformly probabilistically bounded by
\begin{equation}\label{eq_errorBound}
    \text{Pr} \left \{ \left | f(x) - \mu(x, \mathbb{D})  \right |  \leq \eta(x), \forall x\in \mathbb{X} \right \} \ge 1-\delta,
\end{equation}
where $\eta(x) = \sqrt{\beta} \sigma(x) + \gamma$ with 
\begin{align}
    \beta &= 2\sum_{j=1}^{2} \log \left ( \frac{1}{\sqrt{2} \tau } (\bar{x}_j - \underline{x}_j )+1 \right ) -2 \log\delta,  \\
    \gamma &= \left( \sqrt{\beta} L_{\sigma} + L_{f} + L_{\mu}\right)\tau,
\end{align}
in which $\bar{x}_j = \max_{x\in \mathbb{X}} x_j$, $\underline{x}_j = \min_{x\in \mathbb{X}} x_j$ for the $j$-th dimension of $x$. The existence of $\bar{x}_j$ and $\underline{x}_j$ are guaranteed by the compact input domain $\mathbb{X}$. 
The positive constant $L_f$ is the Lipschitz constant for the unknown function, which can be obtained by using the first principle or be approximated as shown in ~\cite{lederer2019uniform}.
The computation of Lipschitz constants $L_{\mu}$ and $L_{\sigma}$ follows
\begin{align}
    &L_{\mu} = L_k \sqrt{M} \big\| (K + \sigma_T^2 I_M)^{-1} Y \big\|, \\
    &L_{\sigma} = \sup_{x,x' \in \mathbb{X}} \left\| \sqrt{\frac{1}{\kappa(x,x) - \kappa(x - x')}} \frac{\kappa(x - x')}{x - x'} \right\|,
\end{align}
where $\kappa(x - x') = \kappa(x,x')$.
The detailed expression of $L_{\mu}$ and $L_{\sigma}$ depends on the choice of the kernel function, which is discussed later in \cref{section_validation}.
 
The theoretical error bound for Gaussian process prediction provides a quantitative way to analyze the performance of a learning-based controller, therefore Gaussian process is recently widely studied in the control area.

\section{Gaussian Process-based control with Optimal Aggregation Strategy}

In this section, the learning-based controller is proposed with an optimal aggregation strategy for distributed GPR models in Section \ref{sec_aggLearning}. The control performance, i.e., the stability, is then analyzed in Section \ref{sec_performanceAnalysis}, in which the tracking error is bounded in a ball area.

\subsection{Design of Learning-based Controller}\label{sec_aggLearning}
Augmenting the training data could lead to improved prediction performance; nevertheless, it necessitates tackling the growing burden of computation and data storage demands~\cite{yangDistributedLearningConsensus2021}. Hence, in light of the growing computational demands and the need for efficient data storage, practical applications widely adopt distributed computation techniques. These techniques involve dividing the complete data set into multiple smaller subsets, which can be effectively organized using a comprehensive aggregation framework.
For GPR, considering $N$ Gaussian process models with different hyperparameters and data sets $\mathbb{D}_i$, $i = 1,2, \cdots, N$. The prediction from individual Gaussian processes varies, i.e., $\mu_i(x) = \mu(x, \mathbb{D}_i)$. To derive the aggregated prediction, popular methods like MOE~\cite{tresp2000mixtures} and GPOE~\cite{cao2014generalized} employ a general framework for distributed GPR~\cite{yangDistributedLearningConsensus2021,dai2023distributed}, which follows a uniform form
\begin{equation}\label{eq_aggEquation}
    \mu(x) = \sum_{i=1}^N w_i \mu_i(x),
\end{equation}
where $w_i$ is the aggregation weight for the $i$-th prediction. The determination of weights in MOE is arbitrary as it lacks a direct and explicit specification. In contrast, GPOE requires significant computation time to determine weights due to the calculation of GPR variance. Therefore,  an optimal aggregation strategy with a focus on control is proposed, aiming to effectively and rationally determine weights to improve the control performance of PMSMs.

In order to track the desired $\varphi_d$, a model-based controller with predicted external torque is designed as follows
\begin{equation} \label{eq_controller}
    I_q \!=\! \frac{2 }{3p\psi } \big( \hat{T} \!+\!J\lambda_1(\varphi \!-\! \varphi_d) \!+\! J\lambda_2 (\omega \!-\! \dot{\varphi}_d) +B\omega  \!+\! J\ddot{{\varphi}_d}  \big).
\end{equation}
Substituting \eqref{eq_controller} into \eqref{eq_system}, the error dynamics of the controlled system becomes
\begin{align}
    \frac{\mathrm{d}}{\mathrm{d} t} \begin{bmatrix}
 \varphi \!-\! \varphi_d\\
\omega  \!-\! \dot{\varphi}_d
\end{bmatrix} \!=\! \begin{bmatrix}
0  & 1\\
\lambda_1  & \lambda_2
\end{bmatrix} \begin{bmatrix}
 \varphi -\varphi_d\\
\omega  -\dot{\varphi}_d
\end{bmatrix} \!+\! \begin{bmatrix}
 0\\
J^{-1}
\end{bmatrix}\left ( T\!-\!\hat{T} \right ) .
\end{align}

Considering the case with perfect information about the external torque, i.e., $\hat{T}=T$, the control gains are designed making matrix $A$ to be Hurwitz with 
\begin{equation}
    A = \begin{bmatrix}
0  & 1\\
\lambda_1  & \lambda_2
\end{bmatrix},
\end{equation}
such that the tracking error $\varphi - \varphi_d$ will asymptotically converge to zero. However, due to the unknown $T$, the compensation $\hat{T}$ is obtained from $N$ distributed Gaussian process models with different data sets. Therefore, the aggregated prediction is 
\begin{align}
    \hat{T} &= \sum_{i=1}^N w_i \hat{f}_i\left (  x_m\right )= \sum_{i=1}^N w_i \mu_i(x_m, \mathbb{D}_i) \nonumber \\
    & = h(x_m)^T w,
\end{align}
where the concatenated prediction is $h(x_m) = [\mu_1(x_m), \cdots, \mu_N(x_m)]^T$ with individual prediction $\mu_i(x_m)$ from the $i$-th GPR using $\mathbb{D}_i$ satisfying $|\mathbb{D}_i| = M_i$ and the lumped aggregation weight is $w = [w_1, \cdots, w_n]^T$ with $\mathbf{1}^T w =1, \forall i = 1, \cdots, N$. 

Therefore, the aggregation weights $w$ are determined by solving the following optimization problem
\begin{align}\label{eq_optimizationProblem}
    &w^* = \underset{w}{\arg\max} ~ e^T P b h(x)^T w, \nonumber \\
    & s.t. \quad \mathbf{1}^T w =1, \nonumber \\
    & ~~~~~~~ w_i \ge 0, \quad \forall i= 1, \cdots, N,
\end{align}
where $b= [0,1]$ and the symmetric positive definite matrix $P$ denote the solution of the continuous algebraic Riccati equation for a given symmetric positive definite $Q$ as
\begin{equation}
    A^TP+PA+Q = 0.
\end{equation}

The existence and uniqueness of $P$ are guaranteed by the designed Hurwitz $A$ by choosing proper gains $\lambda_1$ and $\lambda_2$.

The optimization problem in \eqref{eq_optimizationProblem} has linear objective functions and constraints, forming a constrained linear programming problem, which can be solved in a polynomial time. However, solving \eqref{eq_optimizationProblem} in real-time in the control loop with high frequency is still challenging for the embedded controller. Fortunately, due to the special constraints, an explicit solution exists as 
\begin{equation}
    w_i^* = \begin{cases}
1, \quad \theta^Tg_i = \underset{j}{\min} ~ \theta_j\\
0, \quad  \text{otherwise}
\end{cases},
\end{equation}
where $\theta = h(x)b^T Pe$ and $g_i$ is the $i$-th column of the identical matrix with dimension $N$.

\begin{remark}
    The optimal aggregation weights $w^*$ are not only related to the current states $x$, reflecting the different impacts from distributed data set on one single point, but also to the tracking error $e$.
    Moreover, the choice of $w^*$ considers not only the absolute value of the tracking error $|e|$ but also its direction, which indicates more utilization of the system operation status than the methods in~\cite{yangDistributedLearningConsensus2021,lederer2021gaussian}.
    Intuitively, the optimal weights intend to make the vector $P b h(x)^T w^*$ close to $e$, such that the aggregated prediction induces a small part in $\dot{V}$.
\end{remark}

It is worth noting that the proposed controller and fusion strategy only necessitate the computation of posterior means $\mu_i(x_m)$, which requires $\mathcal{O}(M_i)$ calculations. This approach avoids the computational burden of $\mathcal{O}(M_i^2)$ calculations typically associated with posterior variances $\sigma_i(x_m)$ in most other fusion methods, such as GPOE~\cite{cao2014generalized}. The reduction in computation effort facilitates practical implementation on hardware platforms, particularly for the control of PMSMs. Additionally, apart from the computational considerations, theoretical analysis of control performance is essential and highly significant. The subsequent subsection presents the tracking performance of the proposed learning-based controller with the optimal aggregation strategy.

\subsection{Tracking Performance Analysis}\label{sec_performanceAnalysis}
Stability is an essential point in the control problem of PMSMs, which ensures the system stays in a safe operating area. In this paper, the Lyapunov theory is employed for stability analysis. 

First, a quadratic function with respect to tracking error is selected as a Lyapunov candidate as $V = e^T P e$, with the concatenated tracking error
\begin{align}
    e = \begin{bmatrix}
        \varphi -\varphi_d \\ \omega  -\dot{\varphi}_d
    \end{bmatrix}.
\end{align}
The time derivative of V is written as
\begin{align}\label{eq_Vdot}
    \dot{V} &= e^T \left ( A^T P + PA \right ) e + 2e^TPB\left ( T-\hat{T} \right ) , \nonumber \\
&= - e^T Q e + 2e^TPB\left ( T-h(x_m)^T w \right ),
\end{align}
where $B = J^{-1} b$.
The time derivative of $V$ is considered as a function related to the concatenated aggregation weight $w$, i.e., $\dot{V}(w)$. Choosing the optimal solution $w^*$ from the optimization problem \eqref{eq_optimizationProblem}, the relationship holds
\begin{equation}
    \dot{V}(w^*) \leq \dot{V}(w), \quad \forall w \in \mathbb{R}^N.
\end{equation}

However, due to the unknown $T$, the sign of $\dot{V}(w^*)$ is still not determined. For further analysis, the prediction error bound for Gaussian processes is used as a proxy, such that
\begin{align}\label{eq_Tbound}
    \left \| T - h(x_m)^T w  \right \|  &= \left \| \sum_{i=1}^{N} w_i \left (T-\mu_i(x_m)  \right ) \right \|, \nonumber \\
    &\le \sum_{i=1}^{N} w_i \left \| T- \mu_i(x_m)   \right \| , \nonumber \\
    & \le \sum_{i=1}^{N} w_i \eta_i(x_m),
\end{align}
where the first equality comes from the choice of $w$ with $\mathbf{1}^T w = 1$ and the first inequality follows triangular inequality. Combining with uniform error bound in \eqref{eq_errorBound} and Boole's inequality, the second inequality is derived with a probability of at least $1-N\delta$. Moreover, using the properties of eigenvalues and singular values, $\dot{V}(w)$ is bounded by
\begin{align}
    \dot{V}(w)&\le -\underline{\lambda}(Q)\left \| e \right \|^2 + 2 J^{-1}\left \| e \right \| \left \| P \right \| \left \| T- h(x_m)^T w \right \|, \nonumber \\
    & \le  \underline{\lambda}(Q)\left \| e \right \|^2 + 2 J^{-1}\left \| e \right \| \left \| P \right \| \sum_{i=1}^{N} w_i \eta_i(x_m), 
\end{align}
with the probability of at least $1-N\delta$. The positive part related to $\sum_{i=1}^{N} w_i \eta_i(x)$ impedes the decreasing of the Lyapunov function $V$. Therefore,  the choice of $w$ is considered such that the positive part is as small as possible. To this end, it is assumed that the minimal value of $\sum_{i=1}^{N} w_i \eta_i(x_m)$ achieves at $w=\tilde{w}$ with minimum $\tilde{\eta}(x_m)$ with
\begin{align}
    &\tilde{w} = \underset{w}{\arg\min}\sum_{i=1}^{N} w_i \eta_i(x_m), \\
&\tilde{\eta}(x_m) = \underset{w}{\min}\sum_{i=1}^{N} w_i \eta_i(x_m).
\end{align}
Then, $\dot{V}(\tilde{w})$ is bounded by
\begin{equation}
    \dot{V}(\tilde{w}) \le  \underline{\lambda}(Q)\left \| e \right \|^2 + 2 J^{-1}\left \| e \right \| \left \| P \right \| \tilde{\eta}(x_m),
\end{equation}
whose negativity is achieved when
\begin{equation}
    \left \| e \right \| > 2  \frac{\left \| P \right \|}{J \underline{\lambda}(Q)} \tilde{\eta}(x_m).
\end{equation}
Considering ${V}(\tilde{w})\le \bar{\lambda}(P) \left \| e \right \|^2$, the value of Lyapunov function will converge to
\begin{equation}
    \dot{V}(\tilde{w}) \le 4\frac{\bar{\lambda}(P)}{J^2\lambda^2(Q)}\tilde{\eta}^2_{\max},
\end{equation}
where $\tilde{\eta}^2_{\max} = \sup_{x_m \in \mathbb{X}} \tilde{eta}(x_m)$. Furthermore, as $\dot{V}({w}^*) $ is the minimum of $\dot{V}({w}) $, then it follows with $\dot{V}({w}^*) \le \dot{V}(\tilde{w}) $. Furthermore, according to the property of eigenvalues, i.e.,
\begin{equation}
     \underline{\lambda}(P) \left \| e \right \|^2\le \dot{V}({w}^*) \le 4\frac{\bar{\lambda}(P)}{J^2\lambda^2(Q)}\tilde{\eta}^2_{\max},
\end{equation}
the tracking error will ultimately be bounded by
\begin{equation}
     \left \| e \right \|  \le 2 \sqrt{\frac{\bar{\lambda}(P)}{\underline{\lambda}(P)}} \frac{\|P\|}{J\underline{\lambda}(Q)} \tilde{\eta}_{\max},
\end{equation}
indicating a ball area for tracking error bound.

\begin{remark}
    Considering the positivity of $w_i$ and the relationship $\mathbf{1}^T w = 1$, the weights for control-aware optimal aggregation of experts (COAOE) strategy $\tilde{w}$ are directly chosen as 
    \begin{equation}\label{eq_optimalWeights}
        w_i^* = \begin{cases}
1, \quad \eta_i(x_m) = \underset{j}{\min} ~ \eta_j(x_m)\\
0, \quad  \text{otherwise}
\end{cases},
    \end{equation}
and the corresponding minimum is $\tilde{\eta}(x_m) = \underset{i}{\min}  ~ \eta_i(x_m)$ and $\tilde{\eta}_{\max} = \underset{x_m \in \mathbb{X}}{\sup} \underset{i}{\min} ~  \tilde{\eta}_i(x_m)$, which indicates low computation loads for $\tilde{w}$ and $\tilde{\eta}(x_m)$. However, direct employment of $\tilde{w}$ as the aggregation weights do not necessarily benefit the control performance, since this choice only considers the norm of $f(x_m) - \mu_i(x_m)$ but not its exact value. Despite the best guarantee, it may not provide the best control performance.
\end{remark}

Not similar to other conventional control strategies, such as back-stepping and slide mode control, which achieves asymptotic stability but requires an accurate model or induces vibrated performance, the learning-based controller only achieves an input-to-state-stability (ISS) with the prediction error as the inputs. 
The design of the aggregation weights fully utilizes the potential of the predictions from different Gaussian process models. This designed strategy is proved to be optimal by the given Lyapunov function. 
Moreover, it can also be implemented in the real system due to the low computation load. 

In the next section, the performance of the learning-based controller with the proposed optimal aggregation strategy is demonstrated via simulations.

\section{Validations}
\label{section_validation}
In \cref{subsection_SimulationSetting}, the model of PMSM with given parameters, and the configuration of GPR model are introduced. Then, the performance of the proposed strategy is compared to that of various conventional methods to demonstrate its effectiveness in \cref{subsection_SimulationResult}.

\subsection{Setting} \label{subsection_SimulationSetting}

\begin{figure}[t] 
		\centering
		\def\file{DataSet_Reference.txt}
		\begin{tikzpicture}
			\begin{axis}[xlabel={$\varphi_m$},ylabel={$\omega_m$},
				xmin=-pi, ymin = -1, xmax = pi,ymax=1,legend columns=1,
				width=6cm,height=6cm,legend style={at={(1.05,0.5)},anchor=west},
                clip mode=individual]
				\addplot[only marks, mark=*, mark size=1.5pt, dataPoint1]    table[x = x1 , y  = y1 ]{\file};
                \addplot[only marks, mark=*, mark size=1.5pt, dataPoint2]    table[x = x2 , y  = y2 ]{\file};
                \addplot[only marks, mark=*, mark size=1.5pt, dataPoint3]    table[x = x3 , y  = y3 ]{\file};
                \addplot[only marks, mark=*, mark size=1.5pt, dataPoint4]    table[x = x4 , y  = y4 ]{\file};
				\addplot[black, line width = 0.5mm]    table[x = xr , y  = yr ]{\file};
				\legend{$\mathbb{D}_1$, $\mathbb{D}_2$, $\mathbb{D}_3$, $\mathbb{D}_4$, Reference}
			\end{axis}
		\end{tikzpicture}
        \caption{The distribution of training data sets and the reference trajectory.}
		\label{figure_dataset_reference}
	\end{figure}
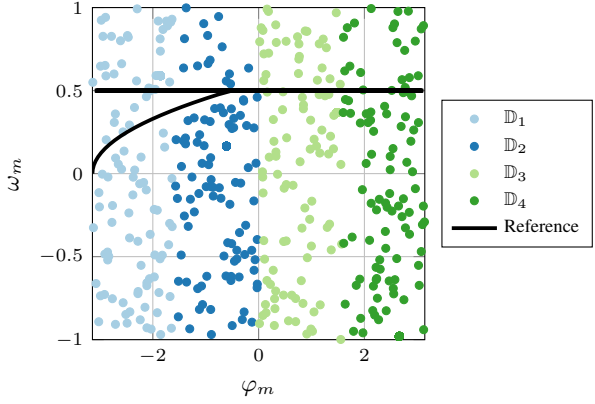

In this validation, a model of PMSM described in \eqref{eq_system} is considered with the parameters as in \cref{table_PMSM_parameter}.
\begin{table}[t]
	\caption{Parameters for the PMSM}
	\begin{center}
		\begin{tabular}{c c c c}
            \hline
			Name & Notation & Value & Unit\\
			\hline
			Moment of inertial & $J$ & $8\times 10^{-5}$ & $\mathrm{kg}/\mathrm{m}^2$\\
			Pole pair & $p$ & $5$ & -- \\
			Damping ratio & $B$ & $0.1$ & -- \\
            Flux linkage & $\psi$ & $0.008$ & $\mathrm{wb}$ \\
			\hline
		\end{tabular}
	\end{center}
	\label{table_PMSM_parameter}
\end{table}
Moreover, the external torque $T$ is considered as
\begin{align}
    T = f(x_m) = 2 \sin (\varphi_m) + 2 \!\times\! 10^{-4} \cos (\varphi_m) \omega_m^2 + 10,
\end{align}
including the constant, position, and velocity-related external torque to simulate the behavior of outside influence, when PMSM is utilized in an integrated starter generator (ISG). Because the PMSM is adopted to ignite and drive the engine to an appropriate speed. The tracking reference is set as
\begin{align}
    \varphi_d (t) = \begin{cases}
        \frac{1}{2} \alpha t^2, & t \in [0,t_{acc}] \\
        \alpha t_{acc} t - \frac{1}{2} \alpha t_{acc}^2, & t \in (t_{acc},\infty)
    \end{cases},
\end{align}
with $t_{acc} = 1$ and $\alpha = 50 \pi /3$, such that the maximal angular velocity achieves $500$ rpm. The control gains $\lambda_1 = -5 \times 10^3$ and $\lambda_2 = -10^4$ are selected for the controller \eqref{eq_controller}. The conversion between the system states $x$ and GPR inputs $x_m$ follows \eqref{eqn_x2xm} with $\upsilon=0.1$ and $\omega_{\mathrm{max}} = 1000$ rpm. The estimated external torque is obtained through $N = 4$ individual GPR models, whose data set covers different input domains as shown in \cref{figure_dataset_reference}, considering different measurement noise, i.e., $\sigma_{T,1} = \sigma_{T,4} = 0.01$  and $\sigma_{T,2} = \sigma_{T,3} = 0.1$. The kernel function for each GPR model is set as identical as
\begin{align}
    \kappa(x_m,x'_m) = \sigma_f^2 \exp \Big(-\frac{\| x_m - x'_m \|^2}{2 l^2} \Big),
\end{align}
with hyperparameters $\sigma_f=1$ and $l=0.2$.

The corresponding Lipschitz constants for the kernel function and posterior variance are written as
\begin{align}
    &L_k = \sigma_f^2 l^{-1} \exp(-0.5) = 5 \exp(-0.5) \\
    &L_\sigma = \sqrt{2} \sigma_f l^{-1} = 5 \sqrt{2}.
\end{align}

In each data set, offline training samples are evenly distributed. The COAOE strategy for distributed GPR follows \eqref{eq_optimalWeights} with $Q$ chosen as the identity matrix. The simulation starts with $t_0 = 0$  and $x_0 = [0,0]^T$, and lasts $10$ seconds.

\subsection{Comparison of Tracking Performance} \label{subsection_SimulationResult}
To demonstrate the advantages of the control performance achieved by the proposed COAOE strategy, a comparative analysis is conducted against other aggregation strategies. The strategies considered for comparison include:
\begin{enumerate}
    \item No prediction: Each GPR model provides the prior mean as the prediction, i.e., $\mu_i(x_m) = m(x_m) = 0$. Then, with any aggregation strategy, $\hat{T} = 0$ holds.
    \item MOE: The aggregation strategy is expressed as 
    \begin{align}
        \mu(x_m) = \sum_{i=1}^N w_i \mu_i(x_m)
    \end{align}
    with $w_i = 1 / N$, $\forall i = 1,\cdots,N$.
    \item GPOE: The aggregation structure is written as
\end{enumerate}
\begin{align}
    &\sigma (x_m) = \sqrt{ \frac{1}{\sum_{i=1}^N w_i \sigma_i^{-2}(x_m)} }, \\
    &\mu(x_m) = \sum_{i=1}^N w_i \frac{\sigma_i^2(x_m)} {\sigma^{2} (x_m)} \mu_i(x_m),
\end{align}
where the aggregation weights are $w_i = 1 / N$, $\forall i = 1,\cdots,N$. 

\begin{figure}[t] 
	\centering
	\begin{tikzpicture}
		\def\file{States.txt}
        \begin{axis}[xlabel={},ylabel={$\varphi$},
			xmin=0.1, ymin = -10, xmax = 9.9,ymax=600,legend columns=3,
			width=0.5\textwidth,height=4.5cm,legend pos= south east,
            xticklabels={,,,}]
			\addplot[black, thick]    table[x = t1 , y  = p1 ]{\file};
            \addplot[greenPlot, thick]    table[x = t2 , y  = p2 ]{\file};
            \addplot[bluePlot, thick]    table[x = t3 , y  = p3 ]{\file};
            \addplot[redPlot, thick]    table[x = t4 , y  = p4 ]{\file};
            \addplot[black, line width = 0.4mm]    table[x = tr , y  = pr ]{\file};
		\end{axis}
		\begin{axis}[xlabel={$t$},ylabel={$\omega$},
			xmin=0.1, ymin = 30, xmax = 9.9,ymax=65,legend columns=3,
			width=0.5\textwidth,height=4.5cm,legend pos= south east,
		  yshift=-3.1cm]
			\addplot[blackPlot, thick]    table[x = t1 , y  = v1 ]{\file};
            \addplot[greenPlot, thick]    table[x = t2 , y  = v2 ]{\file};
            \addplot[bluePlot, thick]    table[x = t3 , y  = v3 ]{\file};
            \addplot[redPlot, thick]    table[x = t4 , y  = v4 ]{\file};
            \addplot[black, line width=0.4mm]    table[x = tr , y  = vr ]{\file};
			\legend{$\!\!$ No prediction $~$, $\!\!$ MOE $~$, $\!\!$ GPOE $~$, $\!\!$ COAOE $~$, $\!\!$ Reference $~$}
		\end{axis}  
	\end{tikzpicture}
	\caption{
		The rotational angle and angular velocity over time.
	}
	\label{figure_states}
\end{figure}
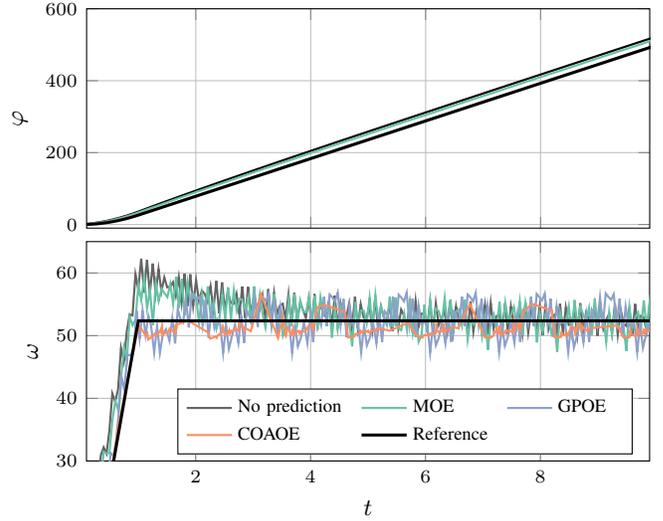

The plots in \ref{figure_states} illustrate the angle and angular velocity of the PMSM under different scenarios. It is evident that the tracking performance is enhanced when utilizing the GPR prediction with consistent control gains. In comparison to the GPOE approach and the proposed COAOE method, the MOE strategy exhibits a more pronounced bias from the reference trajectory. This indicates that the MOE's simplistic aggregation structure performs poorly when employing spontaneous weights. This outcome aligns with intuition, as MOE treats all predictions equally without considering variations in training data accuracy and distribution.

\begin{figure}[t] 
		\centering
		\def\file{Error.txt}
		\begin{tikzpicture}
			\begin{semilogyaxis}[xlabel={$t$},ylabel={$\| e \|$},
			xmin=0.01, ymin = 0.5, xmax = 9.99,ymax=300,legend columns=2,
			width=0.515\textwidth,height=4.5cm,legend pos= north east]
            \addplot[blackPlot, thick]    table[x = t1 , y  = e1 ]{\file};
			\addplot[greenPlot, thick]    table[x = t2 , y  = e2 ]{\file};
			\addplot[bluePlot, thick]    table[x = t3 , y  = e3 ]{\file};
			\addplot[redPlot, thick]    table[x = t4 , y  = e4 ]{\file};
			\legend{$\!\!$ No prediction $~$, $\!\!$ MOE $~$, $\!\!$ GPOE $~$, $\!\!$ COAOE $~$}
		  \end{semilogyaxis}
		\end{tikzpicture}
		\caption{Tracking error over time.}
		\vspace{-0.3cm}
		\label{figure_error}
	\end{figure}
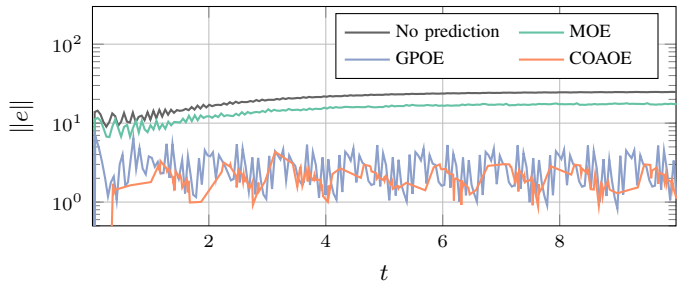

To provide a clearer comparison, the norm of the tracking error over time is depicted in \cref{figure_error}. It shows that the tracking error has less chattering in our method compared to GPOE. These results indicate that while the design of the GPOE structure yields acceptable prediction performance and reduces the tracking error, it is not specifically tailored for the control perspective. As a result, the aggregated prediction may not directly benefit the control process. As the significant computational burden of posterior variance, this limitation hinders the practical application of GPOE in high-frequency PMSM controllers.

In contrast, our proposed method selectively chooses the most useful prediction from a control perspective, leading to smaller tracking errors. The approach circumvents the need for computationally intensive calculations of posterior variance, making it more feasible for implementation in high-frequency PMSM control systems. This phenomenon is also shown in Section \ref{sec_performanceAnalysis} and~\cite{cao2014generalized}, guaranteeing the proposed aggregation strategy is optimal for stability.

\section{Conclusion}
This paper presents a novel control-aware optimal aggregation strategy for tracking control of PMSMs with distributed GPR. The strategy addresses the increasing demand for accurate control in varying and unknown environments by leveraging machine learning techniques, specifically GPR, which offers flexibility in continuous system modeling and guaranteed performance. Based on the Lyapunov stability theory, it eliminates the need for computationally intensive calculations associated with posterior variance. 
By using the proposed controller based on COAOE, the tracking error converges to a bounded error bound ultimately with guarantees. The results indicate that the proposed method outperforms the GPR prediction approach using the MOE and GPOE structures, achieving improved tracking performance with smaller tracking errors.







\bibliographystyle{IEEEtran}
\bibliography{refs}

\begin{thebibliography}{10}
\providecommand{\url}[1]{#1}
\csname url@samestyle\endcsname
\providecommand{\newblock}{\relax}
\providecommand{\bibinfo}[2]{#2}
\providecommand{\BIBentrySTDinterwordspacing}{\spaceskip=0pt\relax}
\providecommand{\BIBentryALTinterwordstretchfactor}{4}
\providecommand{\BIBentryALTinterwordspacing}{\spaceskip=\fontdimen2\font plus
\BIBentryALTinterwordstretchfactor\fontdimen3\font minus
  \fontdimen4\font\relax}
\providecommand{\BIBforeignlanguage}[2]{{%
\expandafter\ifx\csname l@#1\endcsname\relax
\typeout{** WARNING: IEEEtran.bst: No hyphenation pattern has been}%
\typeout{** loaded for the language `#1'. Using the pattern for}%
\typeout{** the default language instead.}%
\else
\language=\csname l@#1\endcsname
\fi
#2}}
\providecommand{\BIBdecl}{\relax}
\BIBdecl

\bibitem{zhang2021coupling}
B.~Zhang, H.-N. Chiu, and Y.-M. Chen, ``{Coupling Network State Equation
  Control for Permanent Magnet Synchronous Motor Emulators},'' in \emph{{2021
  IEEE International Future Energy Electronics Conference (IFEEC)}}.\hskip 1em
  plus 0.5em minus 0.4em\relax IEEE, 2021, pp. 1--5.

\bibitem{yan2018novel}
Z.~Yan, J.~Li, Y.~Wu, and Z.~Yang, ``{A Novel Path Planning for AUV based on
  Objects’ Motion Parameters Predication},'' \emph{{IEEE Access}}, vol.~6,
  pp. 69\,304--69\,320, 2018.

\bibitem{gao2023quasi}
L.~Gao, X.~Dai, M.~Kleeberger, and J.~Fottner, ``{Quasi-static Optimal Control
  Strategy of Lattice Boom Crane based on Large-Scale Flexible Non-linear
  Dynamics},'' in \emph{{Simulation and Modeling Methodologies, Technologies
  and Applications: International Online Conference (SIMULTECH 2021)}}.\hskip
  1em plus 0.5em minus 0.4em\relax Springer, 2023, pp. 153--177.

\bibitem{zhou1998essentials}
K.~Zhou and J.~C. Doyle, \emph{{Essentials of Robust Control}}.\hskip 1em plus
  0.5em minus 0.4em\relax Prentice Hall Upper Saddle River, NJ, 1998, vol. 104.

\bibitem{apte2019disturbance}
A.~Apte, U.~Thakar, and V.~Joshi, ``Disturbance observer based speed control of
  pmsm using fractional order pi controller,'' \emph{IEEE/CAA Journal of
  Automatica Sinica}, vol.~6, no.~1, pp. 316--326, 2019.

\bibitem{bovskovic2020novel}
M.~{\v{C}}. Bo{\v{s}}kovi{\'c}, T.~B. {\v{S}}ekara, and M.~R. Rapai{\'c},
  ``{Novel Tuning Rules for PIDC and PID Load Frequency Controllers Considering
  Robustness and Sensitivity to Measurement Noise},'' \emph{{International
  Journal of Electrical Power \& Energy Systems}}, vol. 114, p. 105416, 2020.

\bibitem{zhao2015adaptive}
S.~Zhao, Y.~Chen, and Y.~Mao, ``Adaptive load observer-based feed-forward
  control in pmsm drive system,'' \emph{Transactions of the Institute of
  Measurement and Control}, vol.~37, no.~3, pp. 414--424, 2015.

\bibitem{yin2022implementation}
Z.~Yin and H.~Zhao, ``{Implementation of Various Neural-Network-Based Adaptive
  Speed PI Controllers for Dual-Three-Phase PMSM},'' in \emph{{IECON 2022--48th
  Annual Conference of the IEEE Industrial Electronics Society}}.\hskip 1em
  plus 0.5em minus 0.4em\relax IEEE, 2022, pp. 1--6.

\bibitem{giles1987learning}
C.~L. Giles and T.~Maxwell, ``Learning, invariance, and generalization in
  high-order neural networks,'' \emph{Applied optics}, vol.~26, no.~23, pp.
  4972--4978, 1987.

\bibitem{williams2006gaussian}
C.~K. Williams and C.~E. Rasmussen, \emph{{Gaussian Processes for Machine
  Learning}}.\hskip 1em plus 0.5em minus 0.4em\relax {MIT Press Cambridge, MA},
  2006, vol.~2, no.~3.

\bibitem{dai2023can}
X.~Dai, A.~Lederer, Z.~Yang, and S.~Hirche, ``{Can Learning Deteriorate
  Control? Analyzing Computational Delays in Gaussian Process-Based
  Event-Triggered Online Learning},'' in \emph{{5th Annual Learning for
  Dynamics {\&} Control Conference}}, 2023.

\bibitem{ignatyev2023sparse}
D.~I. Ignatyev, H.-S. Shin, and A.~Tsourdos, ``{Sparse Online Gaussian Process
  Adaptation for Incremental Backstepping Flight Control},'' \emph{{Aerospace
  Science and Technology}}, vol. 136, p. 108157, 2023.

\bibitem{cao2013efficient}
Y.~Cao, M.~A. Brubaker, D.~J. Fleet, and A.~Hertzmann, ``Efficient optimization
  for sparse gaussian process regression,'' \emph{Advances in Neural
  Information Processing Systems}, vol.~26, 2013.

\bibitem{mutny2018efficient}
M.~Mutny and A.~Krause, ``Efficient high dimensional bayesian optimization with
  additivity and quadrature fourier features,'' \emph{Advances in Neural
  Information Processing Systems}, vol.~31, 2018.

\bibitem{deisenroth2015distributed}
M.~Deisenroth and J.~W. Ng, ``Distributed gaussian processes,'' in
  \emph{International Conference on Machine Learning}.\hskip 1em plus 0.5em
  minus 0.4em\relax PMLR, 2015, pp. 1481--1490.

\bibitem{lederer2022networked}
A.~Lederer, M.~Zhang, S.~Tesfazgi, and S.~Hirche, ``Networked online learning
  for control of safety-critical resource-constrained systems based on gaussian
  processes,'' in \emph{2022 IEEE Conference on Control Technology and
  Applications (CCTA)}.\hskip 1em plus 0.5em minus 0.4em\relax IEEE, 2022, pp.
  1285--1292.

\bibitem{melton2020k}
C.~N. Melton, M.~M. Noack, T.~Ohta, T.~E. Beechem, J.~Robinson, X.~Zhang,
  A.~Bostwick, C.~Jozwiak, R.~J. Koch, P.~H. Zwart \emph{et~al.},
  ``{K-means-driven Gaussian Process Data Collection for Angle-resolved
  Photoemission Spectroscopy},'' \emph{{Machine Learning: Science and
  Technology}}, vol.~1, no.~4, p. 045015, 2020.

\bibitem{lederer2021gaussian}
A.~Lederer, A.~J.~O. Conejo, K.~A. Maier, W.~Xiao, J.~Umlauft, and S.~Hirche,
  ``{Gaussian Process-based Real-Time Learning for Safety Critical
  Applications},'' in \emph{{International Conference on Machine
  Learning}}.\hskip 1em plus 0.5em minus 0.4em\relax PMLR, 2021, pp.
  6055--6064.

\bibitem{tresp2000mixtures}
V.~Tresp, ``{Mixtures of Gaussian Processes},'' \emph{{Advances in Neural
  Information Processing Systems}}, vol.~13, 2000.

\bibitem{cao2014generalized}
Y.~Cao and D.~J. Fleet, ``{Generalized Product of Experts for Automatic and
  Principled Fusion of Gaussian process Predictions},'' \emph{{arXiv preprint
  arXiv:1410.7827}}, 2014.

\bibitem{tresp2000bayesian}
V.~Tresp, ``A bayesian committee machine,'' \emph{Neural computation}, vol.~12,
  no.~11, pp. 2719--2741, 2000.

\bibitem{liu2020gaussian}
H.~Liu, Y.-S. Ong, X.~Shen, and J.~Cai, ``{When Gaussian Process Meets Big
  Data: A Review of Scalable GPs},'' \emph{{IEEE transactions on Neural
  Networks and Learning Systems}}, vol.~31, no.~11, pp. 4405--4423, 2020.

\bibitem{srinivas2012information}
N.~Srinivas, A.~Krause, S.~M. Kakade, and M.~W. Seeger,
  ``{Information-theoretic Regret Bounds for Gaussian Process Optimization in
  the Bandit Setting},'' \emph{{IEEE Transactions on Information Theory}},
  vol.~58, no.~5, pp. 3250--3265, 2012.

\bibitem{lederer2019uniform}
A.~Lederer, J.~Umlauft, and S.~Hirche, ``Uniform error bounds for gaussian
  process regression with application to safe control,'' \emph{Advances in
  Neural Information Processing Systems}, vol.~32, 2019.

\bibitem{yangDistributedLearningConsensus2021}
Z.~Yang, S.~Sosnowski, Q.~Liu, J.~Jiao, A.~Lederer, and S.~Hirche,
  ``{Distributed Learning Consensus Control for Unknown Nonlinear Multi-Agent
  Systems based on Gaussian Processes},'' in \emph{{2021 60th IEEE Conference
  on Decision and Control ({{CDC}})}}.\hskip 1em plus 0.5em minus 0.4em\relax
  {IEEE}, pp. 4406--4411.

\bibitem{dai2023distributed}
X.~Dai, Z.~Yang, M.~Xu, and S.~Hirche, ``{Distributed Event-Triggered Online
  Learning for Multi-Agent System Control using Gaussian Process Regression},''
  \emph{{arXiv preprint arXiv:2304.05138}}, 2023.

\end{thebibliography}

\end{document}